\documentclass[conference]{IEEEtran}
\usepackage{afterpage}

\usepackage[letterpaper, top=0.820in, bottom=1.1in, left=0.667in, right=0.667in, columnsep=0.24in]{geometry}

\usepackage{cite}
\usepackage{amsmath,amssymb,amsfonts}
\usepackage{algorithmic}
\usepackage{graphicx}
\usepackage{tabularx}
\usepackage[T1]{fontenc}
\usepackage{textcomp}
\usepackage{xcolor}
\usepackage{braket} 
\usepackage{tablefootnote}

\usepackage{cite}
\usepackage{amsmath,amssymb,amsfonts}
\usepackage{algorithmic}
\usepackage{graphicx}
\usepackage{textcomp}
\usepackage{xcolor}

\usepackage[letterpaper, top=0.820in, bottom=1.1 in, left=0.667in, right=0.667in, columnsep=0.24in]{geometry}

\begin{document}
\title{Simulation of Quantum Entanglement 
and Quantum Teleportation for Advanced Networks\\
} \author{\IEEEauthorblockN{Sahana Dermal, Asvija Balasubramanyam, Gudapati Naresh Raghava }
\IEEEauthorblockA{sahanad@cdac.in, asvijab@cdac.in, nareshraghava@cdac.in\\
\textit{Quantum Technologies Group} \\
\textit{Centre for Development of 
Advanced Computing (C-DAC)
}\\
Bangalore, Karnataka, India \\
 }
}

\maketitle

\begin{abstract}


This paper introduces the indigenous Quantum Network Simulator developed to simulate various quantum network protocols on classical machines. The paper specifically focuses on the simulation of entanglement generation between two quantum memories using the Barrett-Kok protocol, as well as the teleportation of a single qubit utilizing the produced entangled states. Additionally, the impact of error manipulation and other factors on the fidelity of the teleported states is analyzed, revealing strategies for enhancing teleported state fidelity. The existing few quantum network simulators, allow the simulation of teleportation protocol, but they do not account for the experimental aspects of entanglement generation and distribution, which are crucial for achieving any practical implementation. This work aims to bridge this gap and provide a realistic fidelity of the teleported states.

\end{abstract}


\section{Introduction}

Quantum computing and quantum communications offer unprecedented possibilities in computing and communication realms by leveraging unique quantum mechanics principles like superposition, entanglement, and interference. While quantum computing holds the promise of exponential computational speed-up over classical computing \cite{nielsen:2010:quantum}, quantum communication ensures data security by utilizing quantum cryptography principles \cite{pirandola2020advances}. Quantum technology impacts fields beyond computing and communication, such as drug discovery, defense, and intelligence systems. However, widespread adoption faces significant hurdles due to substantial investment requirements. Quantum network simulators offer a cost-effective alternative for experimentation, eliminating the need for physical networks.

This article talks about the development of a quantum network simulator to simulate quantum network protocols. To evaluate the effectiveness of the software developed, we utilized it to model the creation of entanglement between two terminal nodes in a linear chain quantum network through Barret-Kok protocol \cite{barrett2005efficient}, followed by the teleportation of quantum information using the generated entangled pairs. The simulation results are validated against experimental data \cite{pfaff2014unconditional}, which enhances the reliability and practical relevance of this work. 


There are other quantum network simulators available, such as SeQUeNCe\cite{wu2021sequence} and NetSquid \cite{coopmans2021netsquid} to simulate quantum network protocols. However, these simulators do not account for the experimental aspects of entanglement generation and distribution while simulating teleportation protocol. In the current work proposed in this article, the entanglement generation of quantum states produced using the Barrett-Kok protocol is integrated with the teleportation protocol, incorporating noise parameters to mimic real-world experimental conditions, which are missing in the existing simulators. Entanglement generation is simulated considering its experimental aspects, which, when integrated with teleportation, results in an accurate fidelity of the teleported states. Furthermore, an analysis of factors affecting the fidelity of the teleported states is also conducted. 

The user-friendly Graphical User Interface (GUI) provided in this quantum network simulator makes it easier for users to adjust the parameters and visualize the output. Users can adjust the values of individual parameters, observing their effects through graphs and plots in the output. This interactive feature allows users to gain insights into the extent to which each parameter needs adjustment to achieve a specific value in the output. In essence, it provides a practical means for understanding the relationships between parameter variations and their corresponding impacts on the output data. This is demonstrated in this paper by examining a case study of entanglement generation and teleportation, emphasizing how variations in parameters impact the output data. 

The rest of the paper is organized as follows: Section 2 contains background information, including a detailed explanation of the protocols simulated for this work. Section 3 presents the results and discussions, illustrating the outcomes of each protocol and identifying the parameters that can be adjusted to enhance the fidelity of the teleported state. This is followed by the conclusion in Section 4, which outlines the limitations and potential future work.


\section{Background}

\subsection{Quantum Entanglement and its Generation}
Entanglement, proposed by Einstein, Podolsky, and Rosen in 1935 \cite{einstein1935can}, plays a fundamental role in quantum mechanics, allowing correlation between quantum particles regardless of distance. Various protocols aim to create and distribute entanglement \cite{zhu2015entanglement,qin2018exponentially,gneiting2019disorder}, mainly focusing on photonic networks due to their technological feasibility\cite{semenenko2022entanglement}.

The entanglement generation protocol proposed by Barrett and Kok \cite{barrett2005efficient}, utilizes matter qubits in conjunction with linear optics. Remarkably, the entanglement produced through this protocol claimed to exhibit high fidelity \cite{barrett2005efficient}. Quantum information is stored in quantum memories in the form of matter qubits. Qubit systems can be realized using various platforms such as nitrogen-vacancy centers in diamond, trapped ions with optical transitions, or Pauli-blockade quantum dots with an excess electron \cite{specht2011single,dibos2018atomic,bernien2013heralded}. Every individual quantum memory is embedded within its own optical cavity. Subsequently, the light emitted from these cavities is combined on a 50:50 beam splitter via optical channels. A double-heralded single-photon detection approach is employed to entangle each of these system within the cavity. Quantum memory plays a crucial role in entanglement generation as it serves to store quantum information in the form of matter qubits. In this simulation, single-atom memories, where the spin state of a single atom or ion is stored, are modeled as qubits.

Barett-Kok protocol consists of two rounds. In 
the first round, entanglement between two qubits is generated, while the second round enhances the entanglement and results in a pure maximally entangled state. Initially, both qubits are prepared in the 
$\ket{+}$ state, which can be written as the superposition of spin up and spin down, $\ket{+} = \frac{1}{\sqrt{2}}\left(\ket{\uparrow} + \ket{\downarrow}\right)$. To pump the population in the state $\ket{\downarrow}$ to an excited state $\ket{e}$ an optical $\pi$ pulse is applied to each qubit. The excited photons reach the photon detectors at the Bell-state Measurement (BSM) node by taking a time $t_1$ required to travel from the source to the 
detectors. It is important to make sure that the photons emitted from both directions reach the beam splitter at the same time for ensuring the successful entanglement generation. To achieve this, quantum nodes coordinate timing to stimulate the memories so that it can reach beam splitter simultaneously. After passing through the beamsplitter it get detected at any one of the detectors.

Furthermore, a waiting period of time $t_2$ is implemented to allow the qubit cavity system to relax. After this duration, i.e., when the memory reaches from the excited state to the 
ground state, an X operation is applied to both qubits. This operation results in the flipping of spins from $\ket{\uparrow}$ to $\ket{\downarrow}$ and from $\ket{\downarrow}$ to $\ket{\uparrow}$. Again an optical $\pi$-pulse is applied to each qubit simultaneously to pump the population from the state $\ket{\downarrow}$ to the excited state $\ket{e}$. The excited photons reach the BSM node and get detected at one of the two detectors. The successful completion of two rounds of the protocol results in a pair of entangled memories with maximal entanglement state $\ket{\psi^+} =  \frac{1}{\sqrt{2}} \left(\ket{\uparrow \downarrow } + \ket{\downarrow \uparrow }\right)$ or  $\ket{\psi^-} =  \frac{1}{\sqrt{2}} \left(\ket{\uparrow \downarrow } - \ket{\downarrow \uparrow }\right)$. If the same detector detects a photon in both rounds, it signifies the generation of the state $\ket{\psi^+}$; otherwise, it corresponds to $\ket{\psi^-}$. If no photon detections occur, or if two detections are observed in any round, the procedure is deemed unsuccessful. In such cases, the qubits must be re-prepared for another entanglement generation attempt. Conversely, if only one detector registers a click (precisely one photodetection event) in each round of the procedure, it signifies success. This success leads to the creation of a quantum entangled state. This technique, known as double heralding, demonstrates remarkable resilience against typical experimental errors.

\subsection{Quantum Teleportation}
Quantum teleportation enables the transmission of a quantum state from one location to another. Consider two parties Alice (sender) and Bob (receiver). They are situated at a distance from each other. Suppose Alice wants to send quantum information to Bob, specifically, the qubit state $\ket{\psi} = \alpha \ket{0} + \beta \ket{1}$. This task involves conveying the information pertaining to the values of $\alpha$ and $\beta$ to Bob. The no-cloning theorem dictates that one cannot make an exact copy of an arbitrary unknown quantum state. As a result, Alice is unable to produce an identical copy of the quantum state and transmit it to Bob. Remarkably quantum teleportation enables Alice to transmit the exact quantum state to Bob even if the state is completely unknown to her, by using two classical bits and an entangled qubit pair. This process is referred to as quantum teleportation because, ultimately, Bob will possess the state $\ket{\psi}$, which Alice wanted to send him and Alice will no longer retain it. 

In quantum teleportation, it is necessary to have a shared entangled pair between Alice and Bob, which is obtained here through the Barret-Kok protocol. So there should be two channels established between Alice and Bob: a quantum channel and a classical channel. Alice now possesses a qubit to be teleported $\ket{\psi}$, along with one part of the entangled pair, while Bob holds the other part of the entangled pair $\ket{\phi}$. Alice then performs specific operations 
on her qubits, measures it, and transmits the classical outcomes (measurement results) through a classical communication channel to Bob. Upon receiving the message from Alice, based on the results of the measurement, Bob applies his own set of operations to recover Alice’s qubit on his end.

Step 1: The sender (Alice) applies CNOT gate with the qubit to be teleported $\ket{\psi}$ as control and the part of the entangled pair $\ket{\phi_1}$ as target. Following this, Alice applies a Hadamard gate (H - gate) on the qubit $\ket{\psi}$.

Step 2: Alice performs measurements on both of her qubits and communicates the 
measurement outcomes (two bits of classical information) to Bob.

Step 3: Bob determines which operations to apply to his part of the entangled pair, denoted as $\ket{\phi_2}$, based on the measurement result he received. He may apply an X-gate, Z-gate, I-gate, or both X and Z gates to manipulate $\ket{\phi_2}$ in order to obtain the desired quantum state $\ket{\psi}$.

In an ideal scenario, the teleportation protocol would ensure that the state received by Bob is the same as the states Alice intends to send. However, real-world conditions differ, and this perfect outcome is not achievable. Therefore, in this simulation, errors are introduced deliberately to mimic the conditions of real-world experiments. Currently, various types of errors are incorporated  into the teleportation circuit, including gate errors such as bit flip 
errors, phase flip errors, and stochastic Pauli errors, as well as measurement errors.

The discrepancies observed in the output of the simulation of the teleportation protocol can be attributed to these gate errors and measurement errors. These errors can introduce inaccuracies and disturbances during the transmission and 
measurement processes, leading to deviations between the expected and observed results. Additionally, the fidelity of the entangled pair also plays a crucial role in affecting the quality of the outputs. If the fidelity of the memories involved in the entanglement generation protocol is less than 1, it may result in states that are not entirely entangled, further contributing to errors in the outcomes of the teleportation process degrading the fidelity of the teleported states.


\section{Results and Discussions:}
\subsection{GUI Implementation}
\begin{figure}
    \centering
    \includegraphics[width=\linewidth]{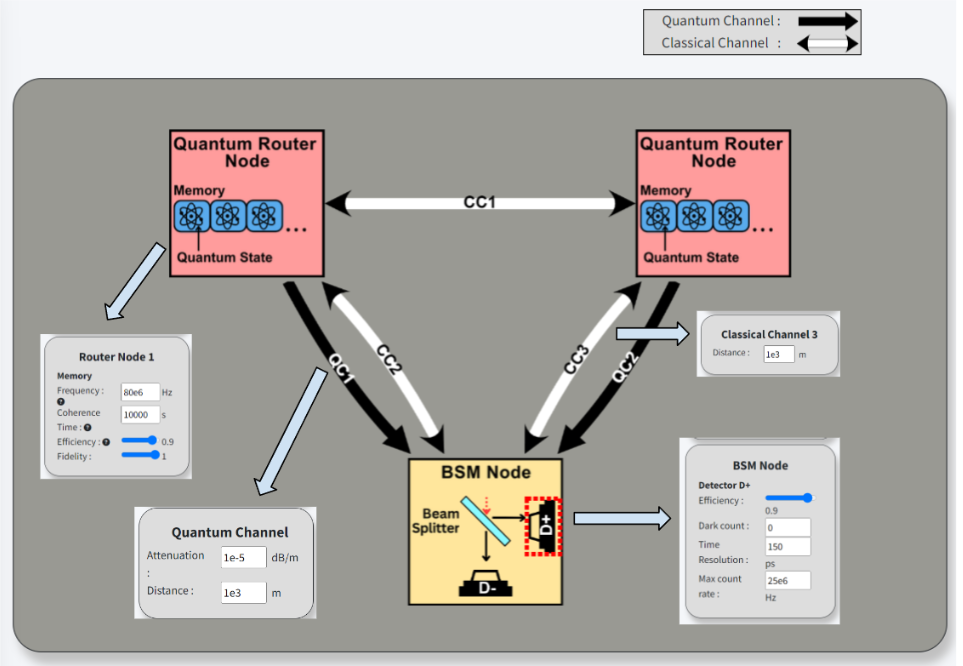}
    \caption{Parameters of each component that user can change in  Barret-Kok entanglement generation protocol.
}
    \label{fig: BK_parameters}
\end{figure}

The successful operation of a software application hinges on the development of an intuitive and user-friendly Graphical User Interface. In the proposed quantum network simulator, users have the flexibility to tailor the characteristics of each physical component integral to quantum communication protocols, such as quantum memory, quantum channel, classical channel, BSM node, and detector. Specifically, within the entanglement generation protocol, users can customize the properties of quantum memories by adjusting parameters such as frequency (frequency of the photon emitted from the cavity), coherence time (lifetime of generated entanglement), efficiency (probability of successful photon emission from the memory), and fidelity. Similarly, parameters related to detectors, including efficiency (probability of accurately detecting an incoming photon), dark count rate (average number of false positive detections per second), time resolution (minimum resolving power of photon arrival time), and maximum count rate (maximum detections per second), can be modified. Additionally, users can experiment with the quantum and classical channels by parameterizing aspects such as channel length (length of the fiber) and attenuation (gradual loss of intensity during transmission through the fiber) as shown in Figure \ref{fig: BK_parameters}, enabling them to observe the resulting changes in outcomes.

\begin{figure}
    \centering
    \includegraphics[width=\linewidth]{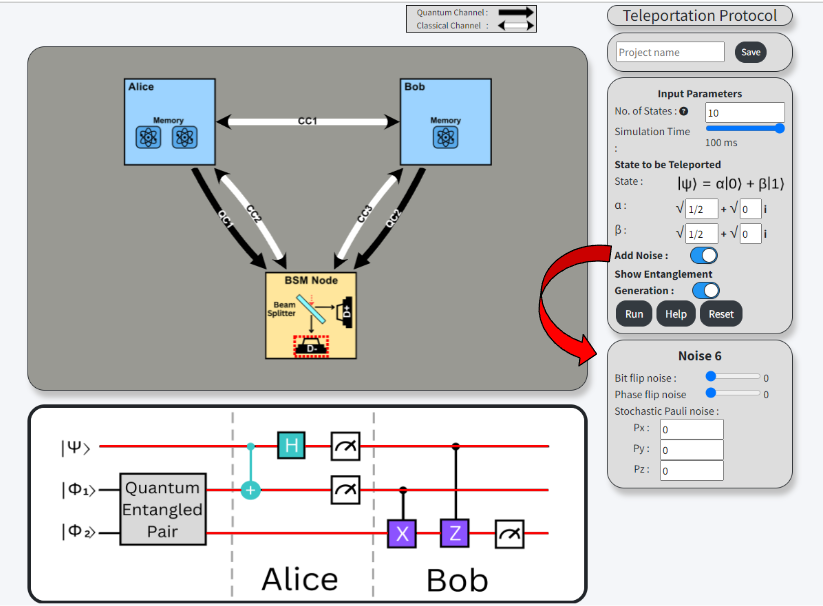}
    \caption{Noise options available in quantum teleportation in the Quantum Network Simulator.
}
    \label{fig:teleportation with noise}
\end{figure}

In quantum teleportation protocol, users have the option to incorporate various types of noise into the circuit. This can be done by selecting specific points on the circuit interface. Users can introduce different types of noises, such as bit flip noise, phase flip noise, and stochastic Pauli noises (Px, Py, Pz) as illustrated in Figure \ref{fig:teleportation with noise} . After running the entanglement generation protocol one can see the time at which each pair of memories got entangled and the number of times the states had to go through round 1 and round 2 to result in an entangled state. Coming to the quantum teleportation protocol in which the entangled states from the Barret-Kok protocol are used, users can see the number of times 0 and 1 measured at the receiver end in the results section which reflects which state got teleported. 

Users can give which state to be teleported and the number of states to be teleported (as shown in Figure \ref{fig:teleportation with noise}) and verify whether the state got teleported perfectly by seeing the result. And can also see how the results are influenced by the effect of the dark count of a detector and the noises in the circuit.

\subsection{Entanglement Generation Protocol and Quantum teleportation}

\begin{figure}
    \centering
    \includegraphics[width=\linewidth]{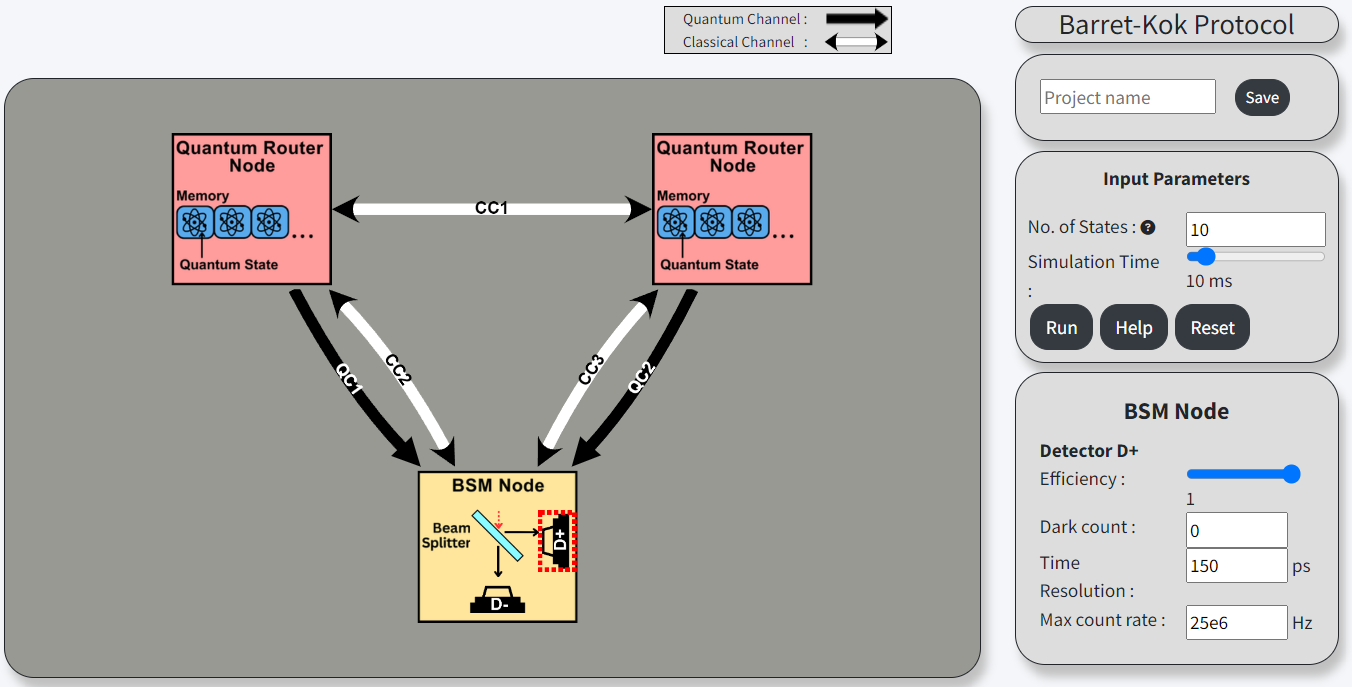}
    \caption{Implementation of Barret-Kok entanglement generation protocol using the Quantum Network Simulator (QNS).
}
    \label{fig:barret-kok}
\end{figure}

\begin{figure}
    \centering
    \includegraphics[width=\linewidth]{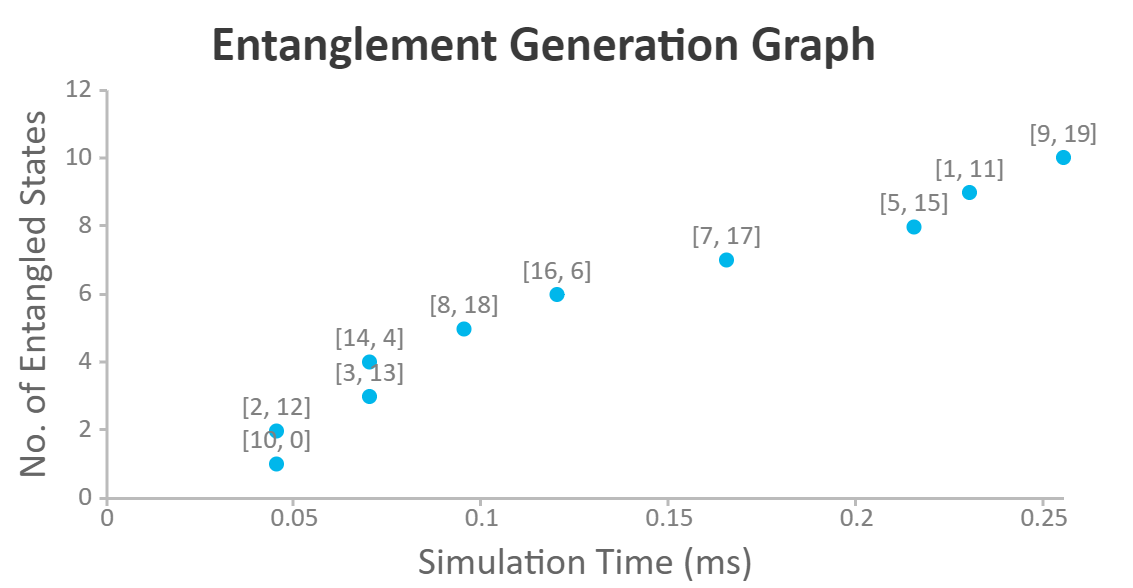}
    \caption{The figure depicts the correlation between the number of entangled states and simulation time. Each data 
point on the graph represents the time at which a pair of memories became entangled.
}
    \label{fig:entanglement_generation}
\end{figure}

\begin{figure}
    \centering
    \includegraphics[width=\linewidth]{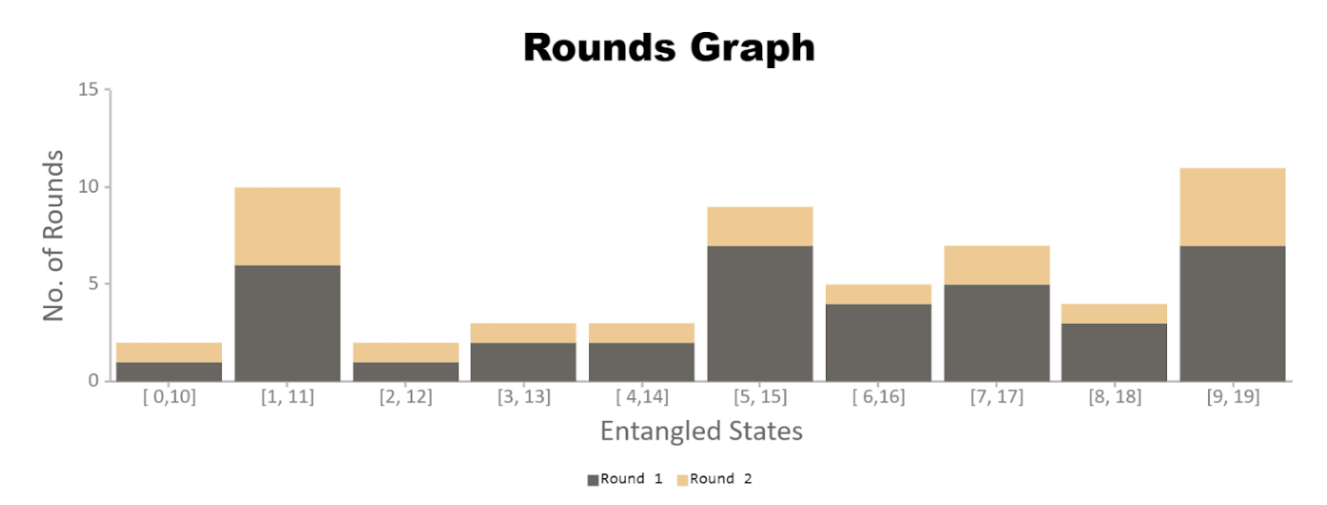}
    \caption{Number of times each state passed through round 1 and round 2, for generating maximally entangled states 
under ideal conditions.
}
    \label{fig:rounds_graph}
\end{figure}

To demonstrate entanglement generation 
based on the Barrett-Kok Protocol, the simulation is initiated (as shown in Figure \ref{fig:barret-kok}) under ideal conditions (i.e., memory 
fidelity = 1, memory efficiency = 1, 
detector efficiency = 1, dark count = 0) to facilitate comprehension. 
Figure \ref{fig:entanglement_generation}  provides information about 
the entangled keys, including the time at 
which each pair became entangled and the 
indices of the memories involved. The increased time required for certain 
memories to become entangled can be 
attributed to the need for multiple attempts 
to achieve successful entanglement 
generation through both rounds 1 and 2 (as 
shown in Figure \ref{fig:rounds_graph}). For instance, the 
entangled pair consisting of memories at 
indices 5 and 15 took longer compared to 
the pair involving memories at indices 2 
and 12. This is because the states at indices 
5 and 15 had to undergo 7 and 2 times, 
respectively, of both rounds 1 and 2, while 
the states at indices 2 and 12 achieved 
maximally entangled states in the first 
attempt itself. In the study, where the comparison is made against experimental findings, memory fidelity is set below 1, and gate errors are incorporated. Similarly, users have the flexibility to modify each parameter and observe its impact on the results through visualizing the corresponding plots generated.


Once the entanglement states are generated 
using the Barrett-Kok protocol, they are 
utilized to teleport a quantum state from 
Alice to Bob. This study takes $\ket{0}$ as the state to be teleported. By manipulating the 
parameters and errors in the Barrett-Kok 
protocol and teleportation circuit, one can 
obtain entangled states and teleported 
qubits with various fidelities. 

In this study, the experimental work 
conducted by Pfaff et al. \cite{pfaff2014unconditional} is simulated, where they 
demonstrated the teleportation of arbitrary 
quantum states between diamond spin 
qubits located 3 meters apart. Their 
achievement of surpassing the classical 
limit with an average state fidelity 
highlights diamond spin qubits as 
promising candidates for realizing quantum 
networks. By fine-tuning the errors at each stage, we achieved states with fidelities that aligned with those observed in the experimental work.
Specifically,  a fidelity of 0.87 
in the entanglement generation step is achieved by 
setting the fidelity of the memories utilized 
in the entanglement generation to 0.87. The 
deterministic Bell-State Measurement 
(BSM) occurring at Alice’s end is a critical 
component of quantum teleportation. 
Similar to the experimental teleportation, 
an average BSM fidelity of 
0.93 is also attained by setting the bit flip error of the 
CNOT gate to  0.018, with the qubit to be 
teleported acting as the control and Alice's 
component of the entangled pair as the 
target. This configuration led to a combined 
probability of applying a bit-flip error on 
both qubits simultaneously of 0.000324.

Maintaining the coherence of the qubit with 
Bob throughout the BSM and feed-forward 
processes are essential. An 
average fidelity of 0.96 for the qubit at 
Bob’s end is attained, similar to the above-mentioned 
experimental setup, by introducing a bit-flip error of 0.09 to that qubit. By fine-tuning error parameters at each stage of the protocol, the required fidelities are achieved culminating in an average fidelity of the teleported states reaching 0.86. This closely aligns with the experimental assertion of achieving a teleportation fidelity of 0.87.

 In addition to comparing the simulation results with experimental data, the impact of certain parameters on the final fidelity of teleported states is also observed. Table \ref{tab:fidelity_table}, illustrates how the fidelity of teleported states varies with the change in errors at different stages of the teleportation circuit and the memory fidelity of entangled states from Barret-Kok Protocol. Here bit-flip errors are introduced during the gate operation at Alice's end, particularly when implementing the CNOT gate. And also these errors are introduced at Bob’s end, where he performs operations (X gate, Z gate, or X and Z gates) based on the measurement outcomes received from Alice.

This analysis offers insights into how 
changes in parameters and errors affect the 
fidelity of the teleported states and provides an understanding of the adjustments required for each parameter to obtain teleported states with a specific value of fidelity.
Specifically, it demonstrates that increasing the memory fidelity of entangled states and reducing gate errors leads to higher fidelity in the teleported states.

\begin{table}
\caption{Factors influencing the fidelity of teleported states}
\begin{tabularx}{\linewidth}{|X|X|X|X|} 
\hline
\begin{tabular}[c]{@{}l@{}}Memory \\ fidelity  of the\\  entangled \\  states\end{tabular} & \begin{tabular}[c]{@{}l@{}}Bit-flip error \\ due to CNOT \\ gate\end{tabular} & \begin{tabular}[c]{@{}l@{}}Bit-flip error \\ due to X gate\end{tabular} & \begin{tabular}[c]{@{}l@{}}Final fidelity \\ of the\\  teleported \\ states\end{tabular} \\ \hline
0.96                                                                               & 0.02                                                                       & 0.02                                                                    & 0.93                                                                               \\ \hline
0.90                                                                               & 0.05                                                                       & 0.05                                                                    & 0.83                                                                               \\ \hline
0.87                                                                              & 0.018                                                                      & 0.09                                                                    & 0.86                                                                              \\ \hline
0.85                                                                               & 0.08                                                                       & 0.08                                                                    & 0.76                                                                               \\ \hline
0.81                                                                               & 0.10                                                                       & 0.10                                                                    & 0.73                                                                               \\ \hline
0.70                                                                               & 0.18                                                                       & 0.18                                                                    & 0.61                                                                               \\ \hline

\end{tabularx}
\label{tab:fidelity_table}
\footnotetext[1]{table footnote 1} 
\end{table}

\section*{Conclusion}

The entanglement generation of two quantum memories and the teleportation of a qubit are simulated using the Quantum Network Simulator (QNS) we developed. The factors affecting the duration of the generation of entangled states and how the fidelity of states at each stage of the teleportation protocol affects the fidelity of the teleported state at the receiver’s end are explored. The simulation results are benchmarked with the experimental research done by Ptaff et al.\cite{pfaff2014unconditional}, in which teleportation using diamond spin qubits with high fidelity is demonstrated. This not only confirms the reliability of the simulation but also emphasizes the crucial role of simulators in examining and validating complex quantum protocols.  The GUI of this quantum network simulator enhances accessibility and usability, by providing users with intuitive controls to manipulate each parameter. This ease of use empowers users to interact with the simulator more effectively, enabling them to explore various scenarios and analyze outcomes with greater depth and precision. These kind of simulators serve as invaluable tools, providing insights that pave the way for real-world applications.

By tweaking errors in the teleportation circuit and adjusting the memory fidelity of entangled states,  how these changes impact the fidelity of teleported states is revealed. Specifically, improving the memory fidelity of entangled states and minimizing gate errors result in higher fidelity in teleported states.  This collaboration between simulation and experimentation	enhances the understanding of quantum phenomena and drives progress in the field of quantum communication and computation. Entanglement generation is a crucial element in quantum networks. As future work, this work can be further extended for use in simulation of  entanglement-based quantum key distribution, quantum repeaters, and other applications.

\bibliographystyle{IEEEtran}
\bibliography{conference_101719}

\end{document}